# Modelling of Inhomogeneous Disk-Loaded Waveguides: Matrix Difference Equations and WKB Approximation

M.I. Ayzatsky[1]

National Science Center Kharkov Institute of Physics and Technology (NSC KIPT),
610108, Kharkov, Ukraine

A new approach to the description of inhomogeneous disk-loaded waveguides (chains of coupled resonators) is proposed. New matrix difference equations based on the technique of coupled integral equations and the decomposition method are obtained. Various approximate approaches have been developed, including the WKB approximation.

## 1 Introduction

Properties of Inhomogeneous Disk-Loaded Waveguides (IDLW) attract the attention of scientists and engineers more than half of century. IDLWs are in fact chains of coupled resonators connected with external waveguides (one or several). There are great number of accelerators and RF sources (and publications) using (describing) such devices. Today, using different computer programs, we can simulate almost any device. But very often for understanding or predicting properties it is necessary to have some equations that can help to interpret the obtained in computer simulation results. Indeed, under computer simulation we obtain the needed results only at the final step of consideration. It is often difficult to develop the effective strategy of choosing the correct values of parameters without preliminary analyses on the base of simplified equations. We can mention here a process of slow-wave electrodynamic structure tuning [1,2,3,4], beam-loading compensation in accelerating structures [5,6,7,8,9,10,11,12], synthesis of the optimum phase velocity distribution along the slow wave structure of TWTs [13,14,15,16,17,18,19,20] and so on.

During the mentioned period two approximate approaches were mainly used to describe IDLWs: a coupled cavity model [21,22,23,24,25,26,27,28,29,30,31] and a smooth waveguide approximation [32,33,34,35,36,37,38]. While the first approach has a mathematical foundation [31], the second, which based on using homogenized parameters, i.e., spatial averaged quantities, relies on physical justification by using the energy conservation law. There are works that study waves in slowly varying band-gap media on the base of analyses of differential operators without assumption that the wavelength is long compared with the size of the repeating cell (see, for example, [39,40,41,42,43] and cited there literature). Results obtained in these works cannot be used for description IDLWs as for IDLWs there are not appropriate smooth differential operators[2]. Taking into account this circumstance it was proposed to use difference equations to study IDLWs [44]. There is possibility that on the base of such approach we can develop a mathematical foundation of a smooth waveguide approximation for IDLWs.

In this work we present new approach for describing IDLW. Using Coupled Integral Equations (CIE) technique (see, for example, [45,46]) and decomposition method [47] we got new matrix difference equations on the basis of which different approximate approaches can be developed, including the WKB approach.

Consideration in the general case is very difficult; therefore, to demonstrate the possibility of the method for describing the IDLV, we choose the simplest case - a chain of cylindrical resonators with annular disks of zero thickness. (the results of studying a homogeneous infinite disk-loaded waveguide with zero-thickness diaphragms see, for example, in [21,22,48,49,50]). As for solution of difference equations it is needed to have additional boundary relationships, we have considered the realistic object – a finite chain of resonators connected with two smooth waveguides (Figure **1**).

---

[1] M.I. Aizatskyi, N.I.Aizatsky; aizatsky@kipt.kharkov.ua
[2] For example, there is not a smooth coordinate transformation to a system in which the perturbed boundary becomes uniform

In Section 2 we give the basic equations. Then we describe the transformation of these equations into a model based on the coupled-integral-equation technique, the main part of which is the coupled matrix difference equations (section 3). In Section 4 we describe numerical implementation of this model. The main results of this work (transformation the coupled matrix difference equations) are presented in Section 5. In Section 6 we present the results of study the homogeneous infinitive DLWs. In Section 7 we give the results of using the proposed approach in studying the properties of IDLWs.

## 2 Chain of the finite number of resonators. Basic equations

Consider a chain of cylindrical resonators with annular discs of zero thickness. The first and last resonators are connected through cylindrical openings to semi-infinite cylindrical waveguides. The geometry of the chain is shown in Figure 1. All resonators are filled with dielectric ($\varepsilon = \varepsilon' + i\varepsilon''$, $\varepsilon'' > 0$). We will consider only axially symmetric fields with $E_z, E_r, H_\varphi$ components (TH). Time dependence is $\exp(-i\omega t)$. In each resonator we expand the electromagnetic field with the waveguide modes

$$\vec{H}^{(k)} = \sum_s \left( h_s^{(k)} \vec{\mathcal{H}}_s^{(k)} + h_{-s}^{(k)} \vec{\mathcal{H}}_{-s}^{(k)} \right), \qquad (1)$$

$$\vec{E}^{(k)} = \sum_s \left( h_s^{(k)} \vec{\mathcal{E}}_s^{(k)} + h_s^{(-k)} \vec{\mathcal{E}}_{-s}^{(k)} \right), \qquad (2)$$

where

$$\mathcal{E}_{s,z}^{(k)} = J_0\left(\frac{\lambda_s}{b_k} r\right) \exp\{\gamma_s^{(k)}(z - z_k)\}, \qquad (3)$$

$$\mathcal{H}_{s,\varphi}^{(k)} = -i\omega \frac{\varepsilon_0 \varepsilon b_k}{\lambda_s} J_1\left(\frac{\lambda_s}{b_k} r\right) \exp\{\gamma_s^{(k)}(z - z_k)\}, \qquad (4)$$

$$\mathcal{E}_{s,r}^{(k)} = -\frac{b_k}{\lambda_s} \gamma_s^{(k)} J_1\left(\frac{\lambda_s}{b_k} r\right) \exp\{\gamma_s^{(k)}(z - z_k)\}, \qquad (5)$$

$$\gamma_s^{(k)2} = \left(\frac{\lambda_s}{b_k}\right)^2 - \frac{\varepsilon \omega^2}{c^2} = \frac{1}{b_k^2}\left(\lambda_s^2 - \frac{\varepsilon' b_k^2 \omega^2}{c^2}\right) - \frac{i\varepsilon'' \omega^2}{c^2}, \qquad (6)$$

$J_0(\lambda_m) = 0$, $z \in [0, d_k]$, $r \in [0, b_k]$.

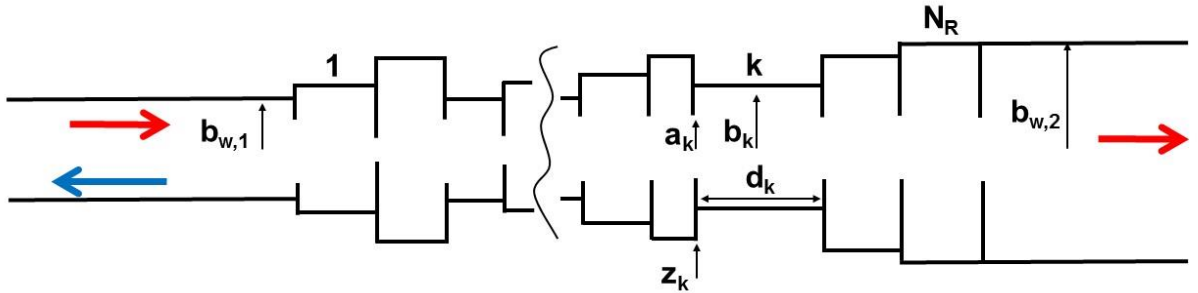

**Figure 1 Chain of resonators with two waveguides**

In the waveguides the electromagnetic field can also be decomposed in terms of TH modes ($k = 1, 2,$)

$$\vec{H}^{(w,k)} = \sum_s \left( G_s^{(k)} \vec{\mathcal{H}}_s^{(w,k)} + G_{-s}^{(k)} \vec{\mathcal{H}}_{-s}^{(w,k)} \right), \tag{7}$$

$$\vec{E}^{(w,k)} = \sum_s \left( G_s^{(k)} \vec{\mathcal{E}}_s^{(w,k)} + G_s^{(-k)} \vec{\mathcal{E}}_{-s}^{(w,k)} \right), \tag{8}$$

where $z_{w,1} = z_1$, $z_{w,2} = z_{N_R+1}$,

$$\mathcal{E}_{s,z}^{(w,k)} = J_0 \left( \frac{\lambda_s}{b_{w,k}} r \right) \exp \left\{ \gamma_s^{(w,k)} (z - z_{w,k}) \right\}, \tag{9}$$

$$\mathcal{H}_{s,\varphi}^{(w,k)} = -i\omega \frac{\varepsilon_0 b_{w,k}}{\lambda_s} J_1 \left( \frac{\lambda_s}{b_{w,k}} r \right) \exp \left\{ \gamma_s^{(w,k)} (z - z_{w,k}) \right\}, \tag{10}$$

$$\mathcal{E}_{s,r}^{(w,k)} = -\frac{b_{w,k}}{\lambda_s} \gamma_s^{(w,k)} J_1 \left( \frac{\lambda_s}{b_{w,k}} r \right) \exp \left\{ \gamma_s^{(w,k)} (z - z_{w,k}) \right\} \tag{11}$$

$$\gamma_s^{(w,k)2} = \frac{1}{b_{w,k}^2} \left( \lambda_s^2 - \frac{b_{w,k}^2 \omega^2}{c^2} \right), \tag{12}$$

The boundary conditions at the interface $z = z_k$ require the continuity of the tangential magnetic fields across the apertures ($k = 2,3,...N_R$)

$$\sum_s \left( h_s^{(k)} \mathcal{H}_{s,\varphi}^{(k)} + h_{-s}^{(k)} \mathcal{H}_{-s,\varphi}^{(k)} \right) = \sum_s \left( h_s^{(k-1)} \mathcal{H}_{s,\varphi}^{(k-1)} + h_{-s}^{(k-1)} \mathcal{H}_{-s,\varphi}^{(k-1)} \right), \ z = z_k, 0 \le r < a_k, \tag{13}$$

or

$$\sum_s \frac{b_k}{\lambda_s} \left( h_s^{(k)} + h_{-s}^{(k)} \right) J_1 \left( \frac{\lambda_s}{b_k} r \right) =$$
$$= \sum_s \frac{b_{k-1}}{\lambda_s} \left( h_s^{(k-1)} \exp\left(\gamma_s^{(k-1)} d_{k-1}\right) + h_{-s}^{(k-1)} \exp\left(\gamma_{-s}^{(k-1)} d_{k-1}\right) \right) J_1 \left( \frac{\lambda_s}{b_{k-1}} r \right), 0 \le r < a_k \tag{14}$$

Multiplying[3] the right and left sides of this relation by a testing function $\psi_{s'}\left( \frac{r}{a_k} \right)$, $s' = 1, 2..., N_m$ and integrating with respect to $r$ from 0 to $a_k$, we get $N_m$ equations

$$\sum_s \frac{b_k}{\lambda_s} R_{s',s}^{\psi(k,1)} \left( h_s^{(k)} + h_{-s}^{(k)} \right) =$$
$$= \sum_s \frac{b_{k-1}}{\lambda_s} R_{s',s}^{\psi(k,2)} \left( h_s^{(k-1)} \exp\left(\gamma_s^{(k-1)} d_{k-1}\right) + h_{-s}^{(k-1)} \exp\left(\gamma_{-s}^{(k-1)} d_{k-1}\right) \right), k = 2,3,...N_R, s' = 1,2...,N_m \tag{15}$$

$$R_{s',s}^{\psi(k,1)} = \int_0^1 \psi_{s'}(x) J_1 \left( \frac{a_k \lambda_s}{b_k} x \right) x dx,$$
$$R_{s',s}^{\psi(k,2)} = \int_0^1 \psi_{s'}(x) J_1 \left( \frac{a_k \lambda_s}{b_{k-1}} x \right) x dx \tag{16}$$

For the first and last resonators we have

$$\sum_s \left( G_s^{(1)} \mathcal{H}_{s,\varphi}^{(w,1)} + G_{-s}^{(1)} \mathcal{H}_{-s,\varphi}^{(w,1)} \right) = \sum_s \left( h_s^{(1)} \mathcal{H}_{s,\varphi}^{(1)} + h_{-s}^{(1)} \mathcal{H}_{-s,\varphi}^{(1)} \right), \ z = z_1, 0 \le r < a_1, \tag{17}$$

$$\sum_s \left( h_s^{(N_R)} \mathcal{H}_{s,\varphi}^{(N_R)} + h_{-s}^{(N_R)} \mathcal{H}_{-s,\varphi}^{(N_R)} \right) = \sum_s G_s^{(2)} \mathcal{H}_{s,\varphi}^{(w,2)}, \ z = z_{N_R+1}, 0 \le r < a_{N_R+1}, \tag{18}$$

Using the same procedure, we get

---

[3] We will use the Moment Method to solve the system of coupled equations]

$$\sum_s \frac{b_{w,1}}{\lambda_s} R^{\psi(w,1)}_{s',s} \left( G^{(1)}_s + G^{(1)}_{-s} \right) = \varepsilon \sum_s \frac{b_1}{\lambda_s} R^{\psi(1,1)}_{s',s} \left( h^{(1)}_s + h^{(1)}_{-s} \right), s' = 1,2...,N_m, \quad (19)$$

$$\sum_s \frac{b_{w,2}}{\lambda_s} R^{\psi(w,2)}_{s',s} G^{(2)}_s =$$
$$\varepsilon \sum_s (-1)^n \frac{b_{N_R}}{\lambda_s} R^{\psi(w,2)}_{s',s} \left( h^{(N_R)}_s \exp\left(\gamma^{(N_R)}_s d_{N_R}\right) + h^{(N_R)}_{-s} \exp\left(\gamma^{(N_R)}_{-s} d_{N_R}\right) \right), s' = 1,2...,N_m \quad (20)$$

The tangential electromagnetic field $E^{(k)}_r(r, z = d_k)$ we expand in terms of a set of basic functions $\varphi_s\left(\frac{r}{a_k}\right)$

$$E^{(k)}_r = \sum_{s=1}^{N_m} C^{(k)}_s \varphi_s\left(\frac{r}{a_k}\right), \quad (21)$$

The boundary condition for electric field at the junction $z = z_k$ can be written as

$$\sum_{s'} \left( h^{(k)}_{s'} \mathcal{E}^{(k)}_{s',r} + h^{(k)}_{-s'} \mathcal{E}^{(k)}_{-s',r} \right) = \begin{cases} \sum_{s'=1}^{N_m} C^{(k)}_{s'} \varphi_{s'}\left(\frac{r}{a_k}\right), & 0 \le r < a_k \\ 0, & a_k \le r < b_k \end{cases},$$

$$\sum_{s'} \left( h^{(k-1)}_{s'} \mathcal{E}^{(k-1)}_{s',r} + h^{(k-1)}_{-s'} \mathcal{E}^{(k-1)}_{-s',r} \right) = \begin{cases} \sum_{s'=1}^{N_m} C^{(k)}_{s'} \varphi_{s'}\left(\frac{r}{a_k}\right), & 0 \le r < a_k \\ 0, & a_k \le r < b_{k-1} \end{cases}. \quad (22)$$

Using the completeness and orthogonality of Bessel functions $J_1\left(\frac{\lambda_s}{b} r\right)$, we obtain ( $k = 1,...,N_R$ )

$$\frac{b_k^3 J_1^2(\lambda_s)}{\lambda_s} \gamma^{(k)}_s h^{(k)}_s sh\left(\gamma^{(k)}_s d_k\right) = -a_{k+1}^2 \sum_{s'=1}^{N_m} R^{\varphi(k+1,2)}_{s,s'} C^{(k+1)}_{s'} + \exp\left(\gamma^{(k)}_{-s} d_k\right) a_k^2 \sum_{s'=1}^{N_m} R^{\varphi(k,1)}_{s,s'} C^{(k)}_{s'}$$
$$\frac{b_k^3 J_1^2(\lambda_s)}{\lambda_s} \gamma^{(k)}_s h^{(k)}_{-s} sh\left(\gamma^{(k)}_s d_k\right) = -a_{k+1}^2 \sum_{s'=1}^{N_m} R^{\varphi(k+1,2)}_{s,s'} C^{(k+1)}_{s'} + \exp\left(\gamma^{(k)}_s d_k\right) a_k^2 \sum_{s'=1}^{N_m} R^{\varphi(k,1)}_{s,s'} C^{(k)}_{s'} \quad (23)$$

where

$$R^{\varphi(k,1)}_{m,s} = \int_0^1 \varphi_s(x) J_1\left(\frac{a_k \lambda_m}{b_k} x\right) x dx,$$
$$R^{\varphi(k+1,2)}_{m,s} = \int_0^1 \varphi_s(x) J_1\left(\frac{a_{k+1} \lambda_m}{b_k} x\right) x dx \quad (24)$$

Consider the case when the dimensions of two waveguides are chosen such that at working frequency $f$ only the dominant mode $TH_{01}$ propagates, and the higher-order modes are all evanescent We will suppose that there is an incident wave that travels from $z = -\infty$ with amplitude $G^{(1)}_1 = 1$ ( $G^{(1)}_s = 0$, $s \ge 2$ ).

Then the boundary condition for electric field at the junction of the first waveguide and the first resonator ( $z = z_1$ ) gives relations

$$-1 + G_{-1}^{(1)} = 2 \frac{a_1^2 \lambda_1}{J_1^2(\lambda_1) b_{w,1}^2 \gamma_1^{(1)} b_{w,1}} \sum_{s'}^{N_m} R_{s',1}^{\varphi(1,1)} C_{s'}^{(1)},$$

$$G_{-s}^{(1)} = 2 \frac{a_1^2 \lambda_s}{J_1^2(\lambda_s) b_{w,1}^2 \gamma_s^{(1)} b_{w,1}} \sum_{s'}^{N_m} R_{s',s}^{\varphi(1,1)} C_{s'}^{(1)}.$$

(25)

Using the same procedure at the junction $z = z_{N_R+1}$, we get

$$G_s^{(2)} = -2 \frac{a_{N_R+1}^2 \lambda_s}{J_1^2(\lambda_s) b_{w,2}^2 \gamma_s^{(2)} b_{w,2}} \sum_{s'}^{N_m} R_{s',s}^{\varphi(N_R+1,2)} C_{s'}^{(N_R+1)}, \ s = 1, 2, \ldots$$

(26)

For the case of one diaphragm between two circular waveguides, we obtain

$$\sum_s \frac{b_{w,1}}{\lambda_s} R_{s',s}^{\psi(w,1)} \left( G_s^{(1)} + G_{-s}^{(1)} \right) = \sum_s \frac{b_{w,2}}{\lambda_s} R_{s',s}^{\psi(w,2)} G_s^{(2)}.$$

(27)

Substitution (25) and (26) into (27) gives such system

$$\sum_{s'}^{N_m} C_{s'}^{(1)} \sum_{s''} \left[ \frac{a_1^2 R_{s,s''}^{\psi(w,1)} R_{s',s''}^{\varphi(1,1)}}{b_{w,1}^2 J_1^2(\lambda_{s''}) \gamma_{s''}^{(1)} b_{w,1}} + \frac{b_{w,2}}{b_{w,1}} \frac{a_1^2 R_{s,s''}^{\psi(w,2)} R_{s',s''}^{\varphi(1,2)}}{b_{w,2}^2 J_1^2(\lambda_{s''}) \gamma_{s''}^{(2)} b_{w,2}} \right] = -\frac{1}{\lambda_1} R_{s,1}^{\psi(w,1)}.$$

(28)

## 3 Coupled Integral Equation Model

Substitution (23) into (15) gives $(N_R - 1)$ systems from $(N_R + 1)$ necessary systems[4] of the Coupled Integral Equation (CIE) technique [45].

$$\sum_{s'=1}^{N_m} T_{s,s'}^{(k,1)} C_{s'}^{(k)} + \sum_{s'=1}^{N_m} T_{s,s'}^{(k,2)} C_{s'}^{(k)} - \sum_{s'=1}^{N_m} T_{s,s'}^{(k,3)} C_{s'}^{(k+1)} - \sum_{s'=1}^{N_m} T_{s,s'}^{(k,4)} C_{s'}^{(k-1)} = 0,$$

$$k = 2, 3, \ldots, N_R, \ s = 1, 2, \ldots, N_m$$

(29)

where

$$T_{s,s'}^{(k,1)} = \frac{a_k}{b_k} \sum_m \frac{\Lambda_m^{(k)}(d_k)}{J_1^2(\lambda_m)} R_{s,m}^{\psi(k,1)} R_{m,s'}^{\varphi(k,1)},$$

$$T_{s,s'}^{(k,2)} = \frac{a_k}{b_{k-1}} \sum_m \frac{\Lambda_m^{(k-1)}(d_{k-1})}{J_1^2(\lambda_m)} R_{s,m}^{\psi(k,2)} R_{m,s'}^{\varphi(k,2)},$$

$$T_{s,s'}^{(k,3)} = \frac{a_{k+1}^2}{a_k b_k} \sum_m \frac{\Lambda_m^{(k)}(0)}{J_1^2(\lambda_m)} R_{s,m}^{\psi(k,1)} R_{m,s'}^{\varphi(k+1,2)},$$

$$T_{s,s'}^{(k,4)} = \frac{a_{k-1}^2}{a_k b_{k-1}} \sum_m \frac{\Lambda_m^{(k-1)}(0)}{J_1^2(\lambda_m)} R_{s,m}^{\psi(k,2)} R_{m,s'}^{\varphi(k-1,1)},$$

(30)

$$\Lambda_m^{(k)}(z) = \frac{ch\left(b_k \gamma_m^{(k)} \frac{z}{d_k} \frac{d_k}{b_k}\right)}{\gamma_m^{(k)} b_k sh(b_k \gamma_m^{(k)} \frac{d_k}{b_k})}.$$

(31)

From (19) and (20) we obtain two additional systems

---

[4] We have $(N_R + 1)$ interfaces

$$\sum_{s'=1}^{N_m} \left( \varepsilon T_{s,s'}^{(1,1)} - W_{s,s'}^{(1)} \right) - \sum_{s'=1}^{N_m} \varepsilon T_{s,s'}^{(1,3)} C_{s'}^{(2)} = \frac{b_{w,1}}{a_1 \lambda_1} R_{s',1}^{\psi(w,1)},$$

$$\sum_{s'=1}^{N_m} \varepsilon T_{s,s'}^{(N_R+1,4)} C_{s'}^{(N_R)} + \sum_{s'=1}^{N_m} \left( W_{s,s'}^{(2)} - \varepsilon T_{s,s'}^{(N_R+1,2)} \right) C_{s'}^{(N_R+1)} = 0$$

(32)

where

$$W_{s',s}^{(1)} = \frac{a_1}{b_{w,1}} \sum_m \frac{R_{s,m}^{\psi(w,1)} R_{s',m}^{\varphi(1,1)}}{J_1^2(\lambda_m) \gamma_m^{(w,1)} b_{w,1}},$$

(33)

$$W_{s',s}^{(2)} = \frac{a_{N_R+1}}{b_{w,2}} \sum_m \frac{R_{s,m}^{\psi(w,2)} R_{s',m}^{\varphi(N_R+1,1)}}{J_1^2(\lambda_m) \gamma_m^{(w,2)} b_{w,2}}.$$

(34)

The reflection and transmission coefficients are given by

$$R_w = G_{-1}^{(1)} = 1 + 2 \frac{a_1^2 \lambda_1}{J_1^2(\lambda_1) b_{w,1}^2 \gamma_1^{(1)} b_{w,1}} \sum_{s'}^{N_m} R_{s',1}^{\varphi(1,1)} C_{s'}^{(1)}$$

(35)

$$T_w = G_1^{(2)} = -2 \frac{a_{N_R+1}^2 \lambda_1}{J_1^2(\lambda_1) b_{w,2}^2 \gamma_1^{(2)} b_{w,2}} \sum_{s'}^{N_m} R_{s',1}^{\varphi(N_R+1,2)} C_{s'}^{(N_R+1)}$$

(36)

Electric field in the $k$-th resonator can be calculated by summing the relevant sequences

$$E_z^{(k)}(z_k, r=0) = \sum_{s=1}^{N_m} T_{s,1}^{E(k)} C_s^{(k)} - \sum_{s''=1}^{N_m} T_{s,2}^{E(k)} C_s^{(k+1)}, \ 0 < z_k < d_k,$$

(37)

where

$$T_{s,1}^{E(k)} = 2 \frac{a_k^2}{b_k^2} \sum_m \frac{\lambda_m R_{m,s}^{\varphi(k,1)} \Lambda_m^{(k)}(z_k - d_k)}{J_1^2(\lambda_m)}$$

$$T_{s,2}^{E(k)} = 2 \frac{a_{k+1}^2}{b_k^2} \sum_m \frac{\lambda_m R_{m,s}^{\varphi(k+1,2)} \Lambda_m^{(k)}(z_k)}{J_1^2(\lambda_m)}$$

(38)

Therefore the set of systems (in matrix form) of the Coupled Integral Equation Model is

$$\left( T^{(k,1)} + T^{(k,2)} \right) C^{(k)} - T^{(k,3)} C^{(k+1)} - T^{(k,4)} C^{(k-1)} = 0, \ k = 2,3,...,N_R,$$

$$\left( \varepsilon T^{(1,1)} - W^{(1)} \right) C^{(1)} - \varepsilon T^{(1,3)} C^{(2)} = \frac{b_{w,1}}{a_1 \lambda_1} R_1^{\psi(w,1)},$$

$$\varepsilon T^{(N_R+1,4)} C^{(N_R)} + \left( W^{(2)} - \varepsilon T^{(N_R+1,2)} \right) C^{(N_R+1)} = 0,$$

(39)

where $T^{(k)}, W$ are $(N_m, N_m)$ complex matrices, $C^{(k)} = \left( C_1^{(k)}, C_2^{(k)}, ..., C_{N_m}^{(N_R+1)} \right)^T$, $R_1^{\psi(w,1)} = \left( R_{1,1}^{\psi(w,1)}, R_{2,1}^{\psi(w,1)}, ..., R_{N_m,1}^{\psi(w,1)} \right)^T$

We can rewrite (39) as

$$T^\Sigma C^\Sigma = R^\Sigma,$$

(40)

where $T^\Sigma$ is a block-tridiagonal matrix, $C^\Sigma = \left( C^{(1)}, C^{(2)}, ..., C^{(N_R+1)} \right)$. Block tridiagonal systems of linear equations are of great interest since they are encountered in a wide variety of problems, in particular, in discrete differential equations (see, for example, [51] and the literature cited there).

## 4 Numerical implementation of the model

In our models we have to make several choices: the kind and the number $N_m$ of the basis functions $\varphi_n$ and the testing functions $\psi_n$, and the upper limit $L_m$ of summation in the sums for calculation of matrix elements $T_{s,s'}$.

In this work we used the entire-domain basis and testing functions. We considered several sets that give analytical expressions for the Hankel transform (24) (coefficients $R^{\varphi(\psi)}$). The simplest case (J-J) is the use of the Bessel functions $\varphi_s(x) = \psi_s(x) = J_1(\lambda_s x)$, $x \in [0,1]$:

$$R_{m',m}^{\psi(k,1)} = R_{m',m}^{\varphi(k,1)} = \int_0^1 \psi_{m'}(x) J_1\left(\frac{a_k \lambda_m}{b_k} x\right) x dx = -\frac{a_k \lambda_m}{b_k} \frac{J_0\left(\frac{a_k \lambda_m}{b_k}\right) J_1(\lambda_{m'})}{\left(\frac{a_k \lambda_m}{b_k}\right)^2 - (\lambda_{m'})^2}$$

$$R_{m',m}^{\psi(k,2)} = R_{m',m}^{\varphi(k,2)} = \int_0^1 \psi_{m'}(x) J_1\left(\frac{a_k \lambda_m}{b_{k-1}} x\right) x dx = -\frac{a_k \lambda_m}{b_{k-1}} \frac{J_0\left(\frac{a_k \lambda_m}{b_{k-1}}\right) J_1(\lambda_{m'})}{\left(\frac{a_k \lambda_m}{b_{k-1}}\right)^2 - (\lambda_{m'})^2}$$

(41)

In this case the edge behavior of electric field is not incorporated into the algorithm. The second case (M-J) is the use the Bessel functions as the testing functions $\psi_s(x) = J_1(\lambda_s x)$, $x \in [0,1]$ and the complete set of functions that fulfil the edge condition on the diaphragm rims as the basis functions (the Meixner basis). We use such Meixner basis [52]

$$\varphi_s(r) = 2\sqrt{\pi} \frac{\Gamma(s+1)}{\Gamma(s-0.5)} \frac{1}{\sqrt{1-r^2}} P_{2s-1}^{-1}\left(\sqrt{1-r^2}\right), \quad (42)$$

where $P_n^m(x)$ are Legendre functions (or spherical functions) of the first kind [53]. The first three functions are:

$$\varphi_1(r) = \frac{r}{\sqrt{1-r^2}}, \quad (43)$$

$$\varphi_2(r) = \frac{r}{\sqrt{1-r^2}} \{-5r^2 + 4\}, \quad (44)$$

$$\varphi_3(r) = \frac{r}{\sqrt{1-r^2}} \{21r^4 - 28r^2 + 8\}, \quad (45)$$

There exist useful integral for our consideration

$$\int_0^1 \varphi_s(t) J_1(xt) t dt = j_{2s-1}(x) = \sqrt{\frac{\pi}{2x}} J_{2s-0.5}(x) \quad (46)$$

$$R_{m',m}^{\varphi(k,1)} = \int_0^1 \varphi_{m'}(x) J_1\left(\frac{a_k \lambda_m}{b_k}\right) x dx = j_{2m'-1}\left(\frac{a_k \lambda_m}{b_k}\right) = \sqrt{\frac{\pi b_k}{2 a_k \lambda_m}} J_{m'-0.5}\left(\frac{a_k \lambda_m}{b_k}\right)$$

$$R_{m',m}^{\varphi(k,1)} = \int_0^1 \varphi_{m'}(x) J_1\left(\frac{a_k \lambda_m}{b_{k-1}}\right) x dx = j_{2m'-1}\left(\frac{a_k \lambda_m}{b_{k-1}}\right) = \sqrt{\frac{\pi b_{k-1}}{2 a_k \lambda_m}} J_{m'-0.5}\left(\frac{a_k \lambda_m}{b_{k-1}}\right)$$

(47)

And the third case (M-M) is the use the Meixner basis as the basis and testing functions.

The simplest geometrical configuration that can give estimations about the "quality" of the chosen sets of functions is the one thin diaphragm in the cylindrical waveguide. Few calculations were performed to obtain the characteristics of the scattering TH waves on the circular diaphragm [54,55,56,57], so we studied the numerical convergence of the results that was obtained with using the Moment Method.

It is known that the Moment Method can lead to ill-conditioned systems of linear equations (see, for example, [58,59,60,61])..The matrix condition number ($M_c$) can ease the correct choice of parameters for obtaining acceptable results.

We considered diffraction of the $TH_{01}$ wave on the circular diaphragm. Results of our calculations are presented in Tables 1-6 (frequency $f = 2.856$ GHz, waveguide radius $b_{w1} = b_{w2} = 4.2$ cm, aperture radius $a = 1.5$ cm). From these results we can make such conclusions:

- there is a wide range of parameters for which the system of linear equations (39) is not ill-conditioned and we can get results with acceptable accuracy;
- using all three sets of functions gives similar results;
- accuracy of J-J sets is worse than M-J and M-M;
- accuracy of M-J sets is the same as M-M sets;
- accuracy of amplitude calculations in the fourth sign and hundredths of a degree in phase is achieved at $N_m$=2 and $L_m$=500 for the M-J and M-M cases.

The correctness of the calculation results is confirmed by comparison with the experimental results (see **Table 7**).

Bellow we will be use the M-J representation with $N_m$=2 and $L_m$=500.

Analysis of more complicated system (see, as example, Table 8, where results of calculation of wave diffraction on the two coupled resonators $b_{w1} = b_{w2} = b_1 = b_2 = 4.2$ cm, aperture radius $a = 1.5$ cm are presented) shows that chosen values of the number of functions give acceptable accuracy of field calculation too.

**Table 1**

| Matrix condition number $M_C$ for the case (J-J) | | | | | | | |
|---|---|---|---|---|---|---|---|
| $L_m \setminus N_m$ | 1 | 2 | 3 | 5 | 10 | 15 | 25 |
| 5 | 1 | 8 | 1065 | 1254795 | | | |
| 10 | 1 | 5 | 13 | 7744 | 1753175 | | |
| 20 | 1 | 5 | 12 | 37 | 106016 | 12079754 | |
| 30 | 1 | 5 | 12 | 36 | 161 | 1026632 | 63851355 |
| 40 | 1 | 5 | 12 | 36 | 152 | 2470 | 4500670 |
| 50 | 1 | 5 | 12 | 36 | 151 | 350 | 63079507 |
| 100 | 1 | 5 | 12 | 36 | 150 | 342 | 964 |
| 500 | 1 | 5 | 12 | 36 | 150 | 341 | 958 |
| 1000 | 1 | 5 | 12 | 36 | 150 | 341 | 958 |
| 1500 | 1 | 5 | 12 | 36 | 150 | 341 | 958 |

**Table 2**

| Module of reflection coefficient \|R\| for the case (J-J) | | | | | | | |
|---|---|---|---|---|---|---|---|
| $L_m \setminus N_m$ | 1 | 2 | 3 | 5 | 10 | 15 | 25 |
| 5 | 0.9179 | 0.8561 | 0.5930 | 0.0000 | | | |
| 10 | 0.9221 | 0.9018 | 0.8922 | 0.7616 | 0.1006 | | |
| 20 | 0.9228 | 0.9038 | 0.8963 | 0.8894 | 0.7677 | 0.4362 | |
| 30 | 0.9229 | 0.9041 | 0.8970 | 0.8908 | 0.8845 | 0.7700 | 0.5985 |
| 40 | 0.9230 | 0.9043 | 0.8972 | 0.8912 | 0.8861 | 0.8773 | 0.6971 |
| 50 | 0.9230 | 0.9043 | 0.8973 | 0.8914 | 0.8866 | 0.8845 | 0.7718 |
| 100 | 0.9230 | 0.9044 | 0.8974 | 0.8916 | 0.8871 | 0.8856 | 0.8842 |
| 500 | 0.9231 | 0.9044 | 0.8974 | 0.8917 | 0.8873 | 0.8858 | 0.8847 |
| 1000 | 0.9231 | 0.9044 | 0.8975 | 0.8917 | 0.8873 | 0.8858 | 0.8847 |
| 1500 | 0.9231 | 0.9044 | 0.8975 | 0.8917 | 0.8873 | 0.8858 | 0.8847 |

**Table 3**

| Phase of reflection coefficient Arg(R) \| for the case (J-J) | | | | | | | |
|---|---|---|---|---|---|---|---|
| $L_m \setminus N_m$ | 1 | 2 | 3 | 5 | 10 | 15 | 25 |
| 5 | -23.38 | -31.12 | -53.63 | -175.08 | | | |
| 10 | -22.76 | -25.61 | -26.85 | -40.40 | -84.16 | | |
| 20 | -22.66 | -25.34 | -26.32 | -27.20 | -39.85 | -63.91 | |
| 30 | -22.64 | -25.29 | -26.24 | -27.03 | -27.81 | -39.65 | -53.44 |
| 40 | -22.63 | -25.27 | -26.21 | -26.98 | -27.61 | -28.68 | -45.69 |
| 50 | -22.63 | -25.27 | -26.20 | -26.95 | -27.55 | -27.81 | -39.49 |
| 100 | -22.62 | -25.26 | -26.18 | -26.92 | -27.48 | -27.68 | -27.85 |
| 500 | -22.62 | -25.25 | -26.18 | -26.91 | -27.46 | -27.65 | -27.79 |
| 1000 | -22.62 | -25.25 | -26.18 | -26.91 | -27.46 | -27.64 | -27.79 |
| 1500 | -22.62 | -25.25 | -26.18 | -26.91 | -27.46 | -27.64 | -27.79 |

**Table 4**

| Matrix condition number $M_C$ for the cases (J-M) (left) and (M-M)(right) | | | | | | | | |
|---|---|---|---|---|---|---|---|---|
| $L_m \setminus N_m$ | 1 | 2 | 3 | 5 | 1 | 2 | 3 | 5 |
| 5 | 1 | 4 | 205 | 167656 | 1 | 4 | 150 | 117270 |
| 10 | 1 | 3 | 7 | 845 | 1 | 3 | 7 | 611 |
| 20 | 1 | 3 | 6 | 28 | 1 | 3 | 5 | 16 |
| 30 | 1 | 3 | 6 | 29 | 1 | 3 | 4 | 10 |
| 40 | 1 | 3 | 6 | 29 | 1 | 2 | 4 | 9 |
| 50 | 1 | 3 | 6 | 28 | 1 | 2 | 4 | 8 |
| 100 | 1 | 3 | 6 | 28 | 1 | 2 | 4 | 7 |
| 500 | 1 | 3 | 6 | 28 | 1 | 2 | 4 | 7 |
| 1000 | 1 | 3 | 6 | 28 | 1 | 2 | 4 | 7 |
| 1500 | 1 | 3 | 6 | 28 | 1 | 2 | 4 | 7 |

**Table 5**

| $L_m \setminus N_m$ | Module of reflection coefficient \|R\| for the cases (J-M) (left) and (M-M)(right) | | | | | | | |
|---|---|---|---|---|---|---|---|---|
| | 1 | 2 | 3 | 5 | 1 | 2 | 3 | 5 |
| 5 | 0.8675 | 0.8364 | 0.5764 | 0.0000 | 0.8366 | 0.8125 | 0.5597 | 0.0000 |
| 10 | 0.8771 | 0.8761 | 0.8718 | 0.7443 | 0.8622 | 0.8577 | 0.8448 | 0.7261 |
| 20 | 0.8797 | 0.8808 | 0.8799 | 0.8779 | 0.8738 | 0.8733 | 0.8715 | 0.8628 |
| 30 | 0.8802 | 0.8817 | 0.8812 | 0.8802 | 0.8767 | 0.8765 | 0.8757 | 0.8726 |
| 40 | 0.8805 | 0.8822 | 0.8819 | 0.8813 | 0.8784 | 0.8784 | 0.8780 | 0.8764 |
| 50 | 0.8806 | 0.8824 | 0.8821 | 0.8817 | 0.8793 | 0.8793 | 0.8790 | 0.8781 |
| 100 | 0.8808 | 0.8827 | 0.8826 | 0.8825 | 0.8812 | 0.8812 | 0.8811 | 0.8809 |
| 500 | 0.8809 | 0.8829 | 0.8829 | 0.8829 | 0.8826 | 0.8826 | 0.8826 | 0.8826 |
| 1000 | 0.8809 | 0.8829 | 0.8829 | 0.8829 | 0.8828 | 0.8827 | 0.8827 | 0.8827 |
| 1500 | 0.8809 | 0.8829 | 0.8829 | 0.8829 | 0.8829 | 0.8828 | 0.8828 | 0.8828 |

**Table 6**

| $L_m \setminus N_m$ | Phase of reflection coefficient Arg(R) for the cases (J-M) (left) and (M-M)(right) | | | | | | | |
|---|---|---|---|---|---|---|---|---|
| | 1 | 2 | 3 | 5 | 1 | 2 | 3 | 5 |
| 5 | -29.84 | -33.24 | -54.80 | 0.00 | -29.84 | -33.24 | -54.80 | 0.00 |
| 10 | -28.70 | -28.82 | -29.34 | -41.90 | -28.70 | -28.82 | -29.34 | -41.90 |
| 20 | -28.40 | -28.26 | -28.37 | -28.61 | -28.40 | -28.26 | -28.37 | -28.61 |
| 30 | -28.34 | -28.15 | -28.21 | -28.34 | -28.34 | -28.15 | -28.21 | -28.34 |
| 40 | -28.30 | -28.09 | -28.13 | -28.20 | -28.30 | -28.09 | -28.13 | -28.20 |
| 50 | -28.29 | -28.07 | -28.10 | -28.15 | -28.29 | -28.07 | -28.10 | -28.15 |
| 100 | -28.26 | -28.03 | -28.04 | -28.05 | -28.26 | -28.03 | -28.04 | -28.05 |
| 500 | -28.24 | -28.00 | -28.01 | -28.01 | -28.24 | -28.00 | -28.01 | -28.01 |
| 1000 | -28.24 | -28.00 | -28.00 | -28.00 | -28.24 | -28.00 | -28.00 | -28.00 |
| 1500 | -28.24 | -28.00 | -28.00 | -28.00 | -28.24 | -28.00 | -28.00 | -28.00 |

**Table 7**

| Comparison of calculation and experimental results | | |
|---|---|---|
| b=1.99 cm, λ=4.5 cm, f=6.662 GHz | | |
| a/b | \|R\| [54] | \|R\| C(M-J) |
| 0.877 | 0.038 | 0.035 |
| 0.745 | 0.174 | 0.176 |
| 0.622 | 0.443 | 0.444 |
| 0.498 | 0.755 | 0.774 |
| 0.445 | 0.845 | 0.874 |

**Table 8**

| Amplitudes (left) and phases (right) of the longitudinal electric field at the center of the second resonator for the case of $TH_{01}$ wave diffraction on the two coupled resonators ($b_{w1} = b_{w2} = b_1 = b_2 = 4.2$ cm, aperture radius $a = 1.5$ cm) ||||||
| --- | --- | --- | --- | --- | --- | --- |
| $L_m \setminus N_m$ | 2 | 3 | 5 | 2 | 3 | 5 |
| 30 | 2.074988 | 2.06417 | 2.042047 | -176.533 | -176.809 | 79808 |
| 40 | 2.089177 | 2.08271 | 2.070325 | -176.247 | -176.411 | -176.725 |
| 50 | 2.087736 | 2.082862 | 2.073635 | -176.135 | -176.26 | -176.49 |
| 100 | 2.094796 | 2.093126 | 2.090004 | -175.913 | -175.957 | -176.034 |
| 500 | 2.103511 | 2.103416 | 2.103136 | -175.802 | -175.808 | -175.815 |
| 1000 | 2.102148 | 2.102152 | 2.102045 | -175.795 | -175.798 | -175.801 |
| 1500 | 2.102413 | 2.102442 | 2.102378 | -175.793 | -175.796 | -175.797 |

## 5 Transformation of the basic equations

We can rewrite the matrix equation (39) as

$$\left(\varepsilon T^{(1,1)} - W^{(1)}\right)C^{(1)} - \varepsilon T^{(1,3)}C^{(2)} = \frac{b_{w,1}}{a_1 \lambda_1} R_1^{\psi(w,1)},$$

$$\left(T^{(2,1)} + T^{(2,2)}\right)C^{(2)} - T^{(2,3)}C^{(3)} - T^{(k,4)}C^{(1)} = 0,$$

$$C^{(k+1)} + \tilde{T}^{(k,4)}C^{(k-1)} = \tilde{T}^{(k)}C^{(k)}, \quad k = 3,4,...,N_R - 1 \quad , \quad (48)$$

$$\left(T^{(N_R,1)} + T^{(N_R,2)}\right)C^{(N_R)} - T^{(N_R,3)}C^{(N_R+1)} - T^{(N_R,4)}C^{(N_R-1)} = 0,$$

$$\varepsilon T^{(N_R+1,4)}C^{(N_R)} + \left(W^{(2)} - \varepsilon T^{(N_R+1,2)}\right)C^{(N_R+1)} = 0,$$

where

$$\begin{aligned} T^{(k)} &= T^{(k,1)} + T^{(k,2)}, \\ \tilde{T}^{(k)} &= T^{(k,3)-1}T^{(k)}, \\ \tilde{T}^{(k,4)} &= T^{(k,3)-1}T^{(k,4)}. \end{aligned} \quad (49)$$

We separated the equations for the first two and the last two resonators from the others, since when the waveguides are matched to the chain, the first and last resonators can be very different from the rest.

WKB asymptotic approximation theory (see [62] and sited there literature) was developed for a class of almost-diagonal ('asymptotically diagonal') linear second-order matrix difference equations

$$C^{(k+2)} + A^{(k)}C^{(k+1)} + B^{(k)}C^{(k)} = 0, \quad (50)$$

by transforming them into the form

$$C^{(k+2)} - 2C^{(k+1)} + C^{(k)} + G^{(k)}C^{(k)} = 0. \quad (51)$$

We shall transform (50) into the other form

$$C^{(k+2)} + C^{(k)} + G^{(k+1)}C^{(k+1)} = 0, \quad (52)$$

We use the procedure similar to that was used in [62].

In equation

$$C^{(k+1)} + \tilde{T}^{(k,4)}C^{(k-1)} = \tilde{T}^{(k)}C^{(k)}, \quad k = 3,4,...,N_R - 1, \quad (53)$$

we put

$$C^{(k)} = \Xi^{(k)}\tilde{C}^{(k)}. \quad (54)$$

with invertible matrices $\Xi^{(k)}$.

Suppose that the matrices $\Xi^{(k)}, k = 2, 3, ..., N_R$ satisfy the equation

$$\Xi^{(k+2)} = \tilde{T}^{(k+1,4)}\Xi^{(k)} \tag{55}$$

with $\Xi^{(2)} = \Xi^{(3)} = I$, $I$ - the unit matrix.

The solution to (55) is

$$\Xi^{(2n)} = \prod_{s=2}^{n} \tilde{T}^{(2s-1,4)}, \qquad \Xi^{(2n+1)} = \prod_{s=2}^{n} \tilde{T}^{(2s,4)}, \quad n = 2, 3, .... \tag{56}$$

The equation (53) takes then the form

$$\tilde{C}^{(k+1)} + \tilde{C}^{(k-1)} = \tilde{\tilde{T}}^{(k)}\tilde{C}^{(k)}, \, k = 3, 4, ..., N_R - 1, \tag{57}$$

where

$$\tilde{\tilde{T}}^{(k)} = \Xi^{(k+1)-1}\tilde{T}^{(k)}\Xi^{(k)}, \tag{58}$$

As the equation (57) is of the second order, we represent the solution of the matrix difference equation (57) as the sum of two new vectors [47]

$$\tilde{C}^{(k)} = \tilde{C}^{(k,1)} + \tilde{C}^{(k,2)}, \, k = 2, 3, ,..., N_R. \tag{59}$$

By introducing new unknowns $\tilde{C}^{(k,i)}$ instead of the one $\tilde{C}^{(k)}$, we can impose an additional condition. This condition we write in the form

$$\tilde{C}^{(k+1)} = M^{(k,1)}\tilde{C}^{(k,1)} + M^{(k,2)}\tilde{C}^{(k,2)}, \, k = 2, 3, ,..., N_R - 1, \tag{60}$$

where $M^{(k,i)}$ are the arbitrary invertible matrices.

Using (59) and (60) we can rewrite (57) as ( $k = 2, 4, ..., N_R - 2$ )

$$
\begin{aligned}
\left(M^{(k+1,1)} - M^{(k+1,2)}\right)\tilde{C}^{(k+1,1)} &= \left\{\left(\tilde{\tilde{T}}^{(k+1)} - M^{(k+1,2)}\right)M^{(k,1)} - I\right\}\tilde{C}^{(k,1)} + \\
&\quad + \left\{\left(\tilde{\tilde{T}}^{(k+1)} - M^{(k+1,2)}\right)M^{(k,2)} - I\right\}\tilde{C}^{(k,2)} \\
\left(M^{(k+1,2)} - M^{(k+1,1)}\right)\tilde{C}^{(k+1,2)} &= \left\{\left(\tilde{\tilde{T}}^{(k+1)} - M^{(k+1,1)}\right)M^{(k,1)} - I\right\}\tilde{C}^{(k,1)} + \\
&\quad + \left\{\left(\tilde{\tilde{T}}^{(k+1)} - M^{(k+1,1)}\right)M^{(k,2)} - I\right\}\tilde{C}^{(k,2)}
\end{aligned}
\tag{61}
$$

Let's choose matrices $M^{(k,i)}$ ($i = 1, 2$) so that they satisfy quadratic matrix equations ( $k = 2, 4, ..., N_R - 2$ )

$$\left(\tilde{\tilde{T}}^{(k+1)} - M^{(k+1,i)}\right)M^{(k+1,i)} = I. \tag{62}$$

It should be noted that these equations do not define $M^{(2,i)}$. As $M^{(k,i)}$ can be chosen arbitrary, we shall take $M^{(2,i)} = M^{(3,i)}$

Then (61) transforms into

$$
\begin{aligned}
M^{(k+1,2)}\left(M^{(k+1,1)} - M^{(k+1,2)}\right)\tilde{C}^{(k+1,1)} &= \left(M^{(k+1,1)} - M^{(k+1,2)}\right)\tilde{C}^{(k,1)} + \\
&\quad + \left(M^{(k,2)} - M^{(k+1,2)}\right)C^{(k,2)} + \left(M^{(k,1)} - M^{(k+1,1)}\right)C^{(k,1)} \\
M^{(k+1,1)}\left(M^{(k+1,2)} - M^{(k+1,1)}\right)\tilde{C}^{(k+1,2)} &= \left(M^{(k+1,2)} - M^{(k+1,1)}\right)\tilde{C}^{(k,2)} + \\
&\quad + \left(M^{(k,1)} - M^{(k+1,1)}\right)\tilde{C}^{(k,1)} + \left(M^{(k,2)} - M^{(k+1,2)}\right)\tilde{C}^{(k,2)}
\end{aligned}
\tag{63}
$$

It can be shown that in our case[5] the matrix $\tilde{\tilde{T}}^{(k)}$ is nondefective, and can be decomposed as

$$\tilde{\tilde{T}}^{(k)} = U^{(k)}\Theta^{(k)}U^{(k)-1}, \tag{64}$$

where $U^{(k)}$ is the matrix of eigen vectors and $\Theta^{(k)} = diag(\theta_1^{(k)}, \theta_2^{(k)}, ..., \theta_{N_m}^{(k)})$, $\theta_i^{(k)}$ - the eigen values.

---

[5] The infinitive uniform disk-loaded waveguide have $2N_m$ different independent solutions (waves). We can expect that this property will be correct for inhomogeneous waveguide too, at least for the case of slowly varying parameters.

Then the solutions of quadratic equations (62) are

$$M^{(k,i)} = U^{(k)}\Lambda^{(k,i)}U^{(k)-1}, \tag{65}$$

where $i = 1, 2$, $\Lambda^{(k,i)} = diag(\lambda_1^{(k,i)}, \lambda_2^{(k,i)}, ..., \lambda_{N_m}^{(k,i)})$ and $\lambda_s^{(k,i)}$ are the solution of the characteristic equations

$$\lambda_s^{(k,i)2} - \theta_s^{(k)}\lambda_s^{(k,i)} + 1 = 0,$$

$$\lambda_s^{(k,1)} = \frac{\theta_s^{(k)}}{2} + \sqrt{\left(\frac{\theta_s^{(k)}}{2}\right)^2 - 1},, \tag{66}$$

$$\lambda_s^{(k,2)} = \frac{\theta_s^{(k)}}{2} - \sqrt{\left(\frac{\theta_s^{(k)}}{2}\right)^2 - 1}$$

The matrices $M^{(k,i)}$ have the same eigen vectors, therefore they are commutative. As $\lambda_s^{(k,1)}\lambda_s^{(k,2)} = 1$, the matrices $M^{(k,i)}$ satisfy the condition

$$M^{(k,1)}M^{(k,2)} = I \tag{67}$$

Using these properties we transform (63) into ($k = 2, 4, ..., N_R - 2$)

$$\tilde{C}^{(k+1,1)} = M^{(k,1)}\tilde{C}^{(k,1)} + \left(\tilde{M}^{(k+1,1)} - I\right)\left(M^{(k,1)} - M^{(k+1,1)}\right)\tilde{C}^{(k,1)} + \tilde{M}^{(k+1,1)}\left(M^{(k,2)} - M^{(k+1,2)}\right)\tilde{C}^{(k,2)}$$
$$\tilde{C}^{(k+1,2)} = M^{(k,2)}\tilde{C}^{(k,2)} + \left(\tilde{M}^{(k+1,2)} - I\right)\left(M^{(k,2)} - M^{(k+1,2)}\right)\tilde{C}^{(k,2)} + \tilde{M}^{(k+1,2)}\left(M^{(k,1)} - M^{(k+1,1)}\right)\tilde{C}^{(k,1)}, \tag{68}$$

where

$$\tilde{M}^{(k,1)} = \left[M^{(k,2)}\left(M^{(k,1)} - M^{(k,2)}\right)\right]^{-1} = U^{(k)-1}\tilde{\Lambda}^{(k,1)}U^{(k)}$$

$$\tilde{M}^{(k,2)} = \left[M^{(k,1)}\left(M^{(k,2)} - M^{(k,1)}\right)\right]^{-1} = U^{(k)-1}\tilde{\Lambda}^{(k,2)}U^{(k)}$$

$$\tilde{\Lambda}^{(k,1)} = diag\left(\frac{1}{1 - \lambda_1^{(k,2)2}}, ..., \frac{1}{1 - \lambda_{N_m}^{(k,2)2}}\right) \tag{69}$$

$$\tilde{\Lambda}^{(k,2)} = diag\left(\frac{1}{1 - \lambda_1^{(k,1)2}}, ..., \frac{1}{1 - \lambda_{N_m}^{(k,1)2}}\right)$$

As

$$\tilde{M}^{(k+1,1)} + \tilde{M}^{(k+1,2)} = I, \tag{70}$$

then from (68) we get

$$\tilde{C}^{(k+1)} = \tilde{C}^{(k+1,1)} + \tilde{C}^{(k+1,2)} = M^{(k,1)}\tilde{C}^{(k,1)} + M^{(k,2)}\tilde{C}^{(k,2)}. \tag{71}$$

It is coincide with our condition (60).

If elements of matrices $M^{(k,i)}$ vary sufficiently slowly with $k$, then the differences $\left|M_{s,m}^{(k+1,i)} - M_{s,m}^{(k,i)}\right|$ are the small values and we can neglect them (Eikonal approximation)

$$\tilde{C}^{(k+1,1)} = M^{(k,1)}\tilde{C}^{(k,1)},$$
$$\tilde{C}^{(k+1,2)} = M^{(k,2)}\tilde{C}^{(k,2)}. \tag{72}$$

If we neglect only nondiagonal terms in (68) we get the WKB approximation (see, [63])

$$\tilde{C}^{(k+1,1)} = \tilde{\tilde{M}}^{(k+1,1)}\tilde{C}^{(k,1)} = \left\{M^{(k+1,1)} + \tilde{M}^{(k+1,1)}\left(M^{(k,1)} - M^{(k+1,1)}\right)\right\}\tilde{C}^{(k,1)},$$
$$\tilde{C}^{(k+1,2)} = \tilde{\tilde{M}}^{(k+1,2)}\tilde{C}^{(k,2)} = \left\{M^{(k+1,2)} + \tilde{M}^{(k+1,2)}\left(M^{(k,2)} - M^{(k+1,2)}\right)\right\}\tilde{C}^{(k,2)}. \tag{73}$$

Finally, we can wright the transformed system

$$\left(\varepsilon T^{(1,1)} - W^{(1)}\right)C^{(1)} - \varepsilon T^{(1,3)}\left(\tilde{C}^{(2,1)} + \tilde{C}^{(2,2)}\right) = \frac{b_{w,1}}{a_1 \lambda_1} R_1^{\psi(w,1)},$$

$$T^{(2)}\left(\tilde{C}^{(2,1)} + \tilde{C}^{(2,2)}\right) - T^{(2,3)}\left(\tilde{C}^{(3,1)} + \tilde{C}^{(3,2)}\right) - T^{(2,4)}C^{(1)} = 0,$$

$k = 2, 3, \ldots, N_R - 2$

$$\tilde{C}^{(k+1,1)} = \left\{M^{(k+1,1)} + \tilde{M}^{(k+1,1)}\left(M^{(k,1)} - M^{(k+1,1)}\right)\right\}\tilde{C}^{(k,1)} + \tilde{M}^{(k+1,1)}\left(M^{(k,2)} - M^{(k+1,2)}\right)\tilde{C}^{(k,2)},$$

$$\tilde{C}^{(k+1,2)} = \left\{M^{(k+1,2)} + \tilde{M}^{(k+1,2)}\left(M^{(k,2)} - M^{(k+1,2)}\right)\right\}\tilde{C}^{(k,2)} + \tilde{M}^{(k+1,2)}\left(M^{(k,1)} - M^{(k+1,1)}\right)\tilde{C}^{(k,1)},$$

$$T^{(N_R)}\Xi^{(N_R)}\left(M^{(N_R-1,1)}\tilde{C}^{(N_R-1,1)} + M^{(N_R-1,2)}\tilde{C}^{(N_R-1,2)}\right) -$$

$$-T^{(N_R,4)}\Xi^{(N_R-1)}\left(\tilde{C}^{(N_R-1,1)} + \tilde{C}^{(N_R-1,2)}\right) - T^{(N_R,3)}C^{(N_R+1)} = 0,$$

$$\varepsilon T^{(N_R+1,4)}\Xi^{(N_R)}\left(M^{(N_R-1,1)}\tilde{C}^{(N_R-1,1)} + M^{(N_R-1,2)}\tilde{C}^{(N_R-1,2)}\right) + \left(W^{(2)} - \varepsilon T^{(N_R+1,2)}\right)C^{(N_R+1)} = 0.$$

(74)

The electric field in the $k$-th resonator is determined by the vectors $C^{(k)}$ and $C^{(k+1)}$ (see, (37)) which can be calculated by the formula

$$C^{(k)} = C^{(k,1)} + C^{(k,2)} = \Xi^{(k)}\tilde{C}^{(k,1)} + \Xi^{(k)}\tilde{C}^{(k,2)} \tag{75}$$

## 6 Infinite Homogeneous DLW

If we omit the presence of boundaries for the uniform chain of resonators ($b_k = b$, $a_k = a$), we obtain the equations describing an infinite homogeneous disk-loaded waveguide

$$\tilde{C}^{(k+1,1)} = M^{(1)}\tilde{C}^{(k,1)}$$
$$\tilde{C}^{(k+1,2)} = M^{(2)}\tilde{C}^{(k,2)}. \tag{76}$$

For the uniform chain of resonators matrices $\Xi^{(k)} = I$ and $C^{(k,i)} = \tilde{C}^{(k,i)}$.

It can be shown that the general solutions of the difference matrix equations (76) are

$$C^{(k,i)} = \sum_{s=1}^{N_m} B_s^{(i)} \lambda_s^{(i)k} U_s. \tag{77}$$

where $B_s^{(i)}$ - are constants, $\lambda_s$ (characteristic or Floquet multiplier) are the solutions of the characteristic equations

$$\lambda_s^2 - \theta_s \lambda_s + 1 = 0, \tag{78}$$

with $\theta_s$ and $U_s$ that are eigen values and eigen vectors of matrix $\tilde{T}$

$$\tilde{T}U_s = \mu_s U_s. \tag{79}$$

From (78) it follows that

$$\lambda_{s,1}\lambda_{s,2} = 1. \tag{80}$$

This property of the Floquet multipliers (along with the assumption that $\varepsilon'' \neq 0$) guarantees that problem (40) is well-conditioned, at least in the case when matrix elements are slowly changing [64].

Analysis of the solution (77) shows that representation (59) is not a trivial decomposition into forward and backward waves. Decomposition (59) with the conditions (60) and (62) divide the solution of matrix difference equation (57) into two parts each of which is generalization of concepts forward and backward waves, especially in the case of inhomogeneous waveguides.

Starting from the $N_m$-dimensional system, in the case of homogeneous waveguide we can obtain the difference equation that describes the behavior of one component of the vectors $C^{(k)}$, say $C_1^{(k)}$. The above solution of equation for $C^{(k)}$ (77) shows that the characteristic equation of this difference equation must have roots that coincide with the $2N_m$ eigenvalues $\lambda_s$.

The general form of this equation can be prompted by considering the simplest case $N_m = 2$ in system (39) for infinite chain

$$\begin{aligned}\left(C_1^{(k+1)} + C_1^{(k-1)}\right) - \tilde{T}_{1,1}C_1^{(k)} &= \tilde{T}_{1,2}C_2^{(k)} \\ \left(C_2^{(k+1)} + C_2^{(k-1)}\right) - \tilde{T}_{2,2}C_2^{(k)} &= \tilde{T}_{2,1}C_1^{(k)} \end{aligned} \quad (81)$$

We introduce the commutative operators $\hat{L}_i$

$$\hat{L}_i = \hat{\sigma}^+ + \hat{\sigma}^- - \tilde{T}_{i,i}, \quad (82)$$

where $\hat{\sigma}^+ \left(\hat{\sigma}^+ b^{(k)} = b^{(k+1)}\right)$ and $\hat{\sigma}^- \left(\hat{\sigma}^- b^{(k)} = b^{(k-1)}\right)$ - are shift operators. From (81) we can get such equation

$$\widehat{\det} \begin{pmatrix} \hat{L}_1 & -\tilde{T}_{1,2} \\ -\tilde{T}_{2,1} & \hat{L}_2 \end{pmatrix} C_1^{(k)} = 0, \quad (83)$$

where the operator $\widehat{\det}$ is defined on the base of rules of common determinants[6]

$$\widehat{\det} \begin{pmatrix} \hat{L}_1 & -\tilde{T}_{1,2} \\ -\tilde{T}_{2,1} & \hat{L}_2 \end{pmatrix} = \hat{L}_1 \hat{L}_2 - \tilde{T}_{1,2}\tilde{T}_{2,1}, \quad (84)$$

It can be shown that in general case we get the equation

$$\widehat{\det} \begin{pmatrix} \hat{L}_1 & -T_{1,2} & \dots & -T_{1,N_m} \\ -T_{2,1} & \hat{L}_2 & \dots & \dots \\ \dots & \dots & \dots & \dots \\ -T_{N_m,1} & -T_{N_m,1=2} & \dots & \hat{L}_{N_m} \end{pmatrix} C_1^{(k)} = 0, \quad (85)$$

with the characteristic equation

$$\begin{vmatrix} \lambda^2 - T_{1,1}\lambda + 1 & -\lambda T_{1,2} & \dots & -\lambda T_{1,N_m} \\ -\lambda T_{2,1} & \lambda^2 - T_{2,2}\lambda + 1 & \dots & \dots \\ \dots & \dots & \dots & \dots \\ -\lambda T_{N_m,1} & -\lambda T_{N_m,1=2} & \dots & \lambda^2 - T_{N_m,N_m}\lambda + 1 \end{vmatrix} = 0. \quad (86)$$

Numerical calculations show that (78) and (86) give the values of $\lambda_s$ which coincide with good accuracy.

It can be shown that equations for other components of vector $C^{(k)}$ are the same as (85).

Values of several first characteristic multipliers are given in Table 9 and Table 10 ($a = 1.5$ cm, $b = 4,23189$ cm, $d = 3.4989$ cm). These values are fitted well the existing theories of periodic waveguides. For the frequency that lay in the first propagation zone we have two propagation waves and the infinitive number of evanescent waves with characteristic multipliers with zero phases (see Table 9 and Table 10). Between the first and the second propagation zones there are only evanescent oscillations with the phase shift per cell equals 180 degree. The other evanescent oscillations have zero phase shift (see Table 10). For considered above cases of homogeneous resonator chains evanescent waves play important roles only at the ends of chains as they decrease very fast from the inhomogeneity (see values of characteristic multipliers of evanescent oscillations in Table 9 and Table 10).

Analysis of the results from Table 9 and Table 10 also show that choice of the values of $N_m = 2$ and $L_m = 500$ give good accuracy for dispersive characteristics too.

---

[6] We have deal with commutative matrices

**Table 9**

| Amplitudes and phases of the Floquet multipliers for different $N_m$, $f = 2.856$ GHz, $L_m = 500$ | | | | | |
|---|---|---|---|---|---|
| $\|\lambda_s\|$ | $Arg\,\lambda_s,°$ | $\|\lambda_s\|$ | $Arg\,\lambda_s,°$ | $\|\lambda_s\|$ | $Arg\,\lambda_s,°$ |
| $N_m = 1$ | $N_m = 1$ | $N_m = 2$ | $N_m = 2$ | $N_m = 5$ | $N_m = 5$ |
| | | | | 5.03E+09 | 0.00E+00 |
| | | | | 6.79E+07 | 0.00E+00 |
| | | | | 4.44E+05 | 0.00E+00 |
| | | 2.47E+03 | 0.00E+00 | 2.66E+03 | 0.00E+00 |
| 1.00E+00 | -1.19E+02 | 1.00E+00 | -1.20E+02 | 1.00E+00 | -1.20E+02 |
| 1.00E+00 | 1.19E+02 | 1.00E+00 | 1.20E+02 | 1.00E+00 | -1.20E+02 |
| | | 4.06E-04 | 0.00E+00 | 3.76E-04 | 0.00E+00 |
| | | | | 2.25E-06 | 0.00E+00 |
| | | | | 1.47E-08 | 0.00E+00 |
| | | | | 1.99E-10 | 0.00E+00 |

**Table 10**

| Amplitudes and phases of the Floquet multipliers for different $N_m$, $f = 4$ GHz, $L_m = 500$ | | | | | |
|---|---|---|---|---|---|
| $\|\lambda_s\|$ | $Arg\,\lambda_s,°$ | $\|\lambda_s\|$ | $Arg\,\lambda_s,°$ | $\|\lambda_s\|$ | $Arg\,\lambda_s,°$ |
| $N_m = 1$ | $N_m = 1$ | $N_m = 2$ | $N_m = 2$ | $N_m = 5$ | $N_m = 5$ |
| | | | | 4.22E+09 | 0.00E+00 |
| | | | | 5.28E+07 | 0.00E+00 |
| | | | | 3.10E+05 | 0.00E+00 |
| | | 1.22E+03 | 0.00E+00 | 1.30E+03 | 0.00E+00 |
| 1.74E+01 | 1.80E+02 | 1.69E+01 | 1.80E+02 | 1.69E+01 | 180 |
| 5.74E-02 | 1.80E+02 | 5.91E-02 | 1.80E+02 | 5.90E-02 | 180 |
| | | 8.22E-04 | 0.00E+00 | 7.72E-04 | 0.00E+00 |
| | | | | 3.23E-06 | 0.00E+00 |
| | | | | 1.89E-08 | 0.00E+00 |
| | | | | 2.37E-10 | 0.00E+00 |

## 6 Finite Chain of Resonators

We wrote two computer codes. The first is based on the system (39), the second – on the transformed system (74). All results that are given bellow were calculated with $N_m = 2^7$, $L_m = 500$.

These codes give practically the same results. It is confirmed by results of calculation that are presented in Figure 2 ($\varepsilon = 1$[8]). There differences between amplitudes of electric fields at the centers of resonators calculated on the base of systems (39) and (74) for homogeneous and inhomogeneous waveguides with 60 resonators are given ($d_{1-60} = 3.4989$ cm, $b_{2-59} = 4.16595$ cm, $b_1 = b_{60} = 4.19825$ cm, changes in the size of the apertures are shown in Figure 3, $a_1 = a_{61} = 1.7661$ cm, $f = 2.856$ GHz). Here and below we consider the propagation of an incident $TH_{01}$ wave with a unit amplitude through the IDLW, shown schematically in Figure **1**. At the selected frequency and for a homogeneous DLW with $a_{2-59} = 1.3$ cm (the first disk aperture distribution (1) in Figure 3) the phase shift per cell[9] in the DLW equals $2\pi/3$ and the reflection coefficient is $R = $ 7.86E-04 ($T = 0.9999$).

---

[7] Taking such value we include in consideration one propagation wave and one evanescent oscillation

[8] If $\text{Im}\,\varepsilon \neq 0$ the differences become greater, but less than 1.E-6

Consider the accuracy of WKB approximation in the case of IDLW with the geometric dimensions indicated above (the second disk aperture distribution (2) in Figure 3). Parameters of this IDLW change along the waveguide at a moderate gradient. Results of comparison of electric field distributions calculated on the base of systems (39) and (73) are presented in Figure 4 and Figure 5 (1). We also present a comparison of the electric field distributions calculated on the base of systems (39) and (72) (Figure 4 and Figure 5 – (2))

Results of comparisons show that for moderate gradient of IDLW parameters the WKB approximation gives suitable accuracy, while the results of the Eikonal approximation differ from the exact ones more significantly, especially in the phase distribution.

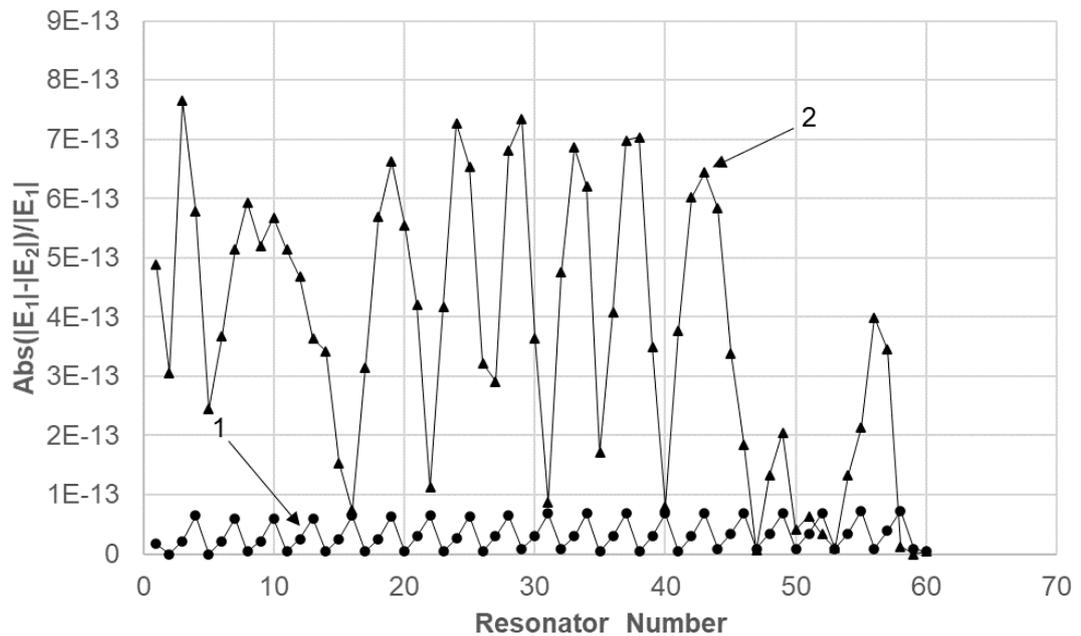

**Figure 2 Differences between amplitudes of electric fields at the centers of resonators calculated on the base of systems (39) and (74) for two DLW: homogeneous (1) and inhomogeneous (2)**

---

[9] In the first propagation zone

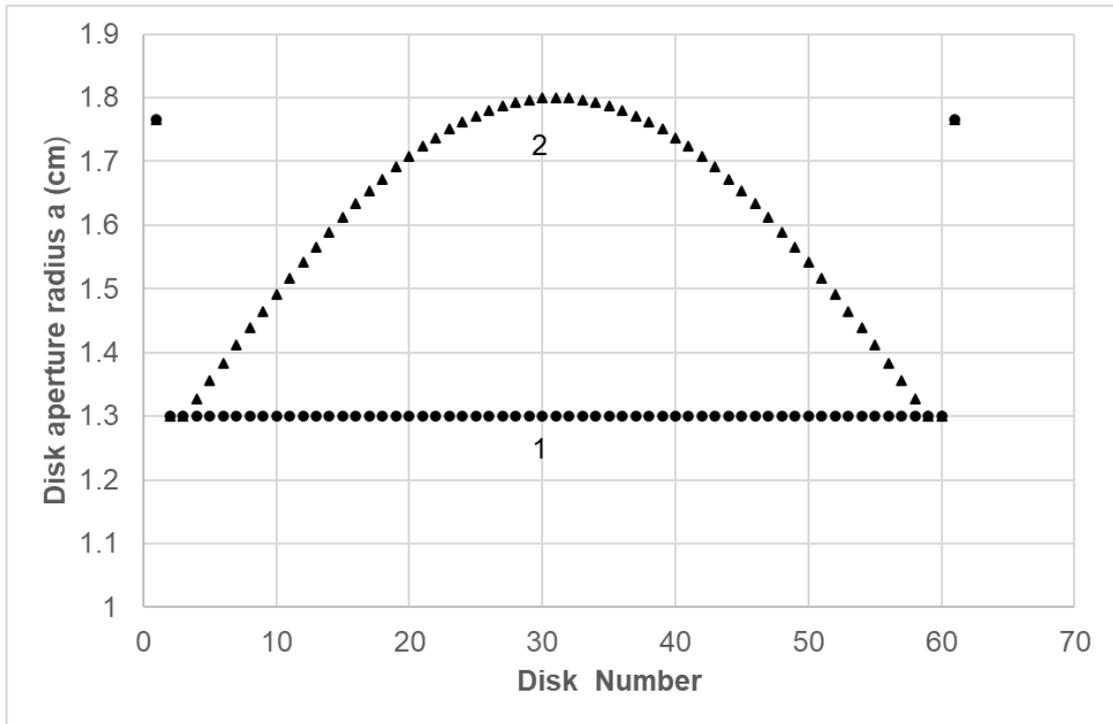

**Figure 3 Dimensions of the apertures considered waveguides**

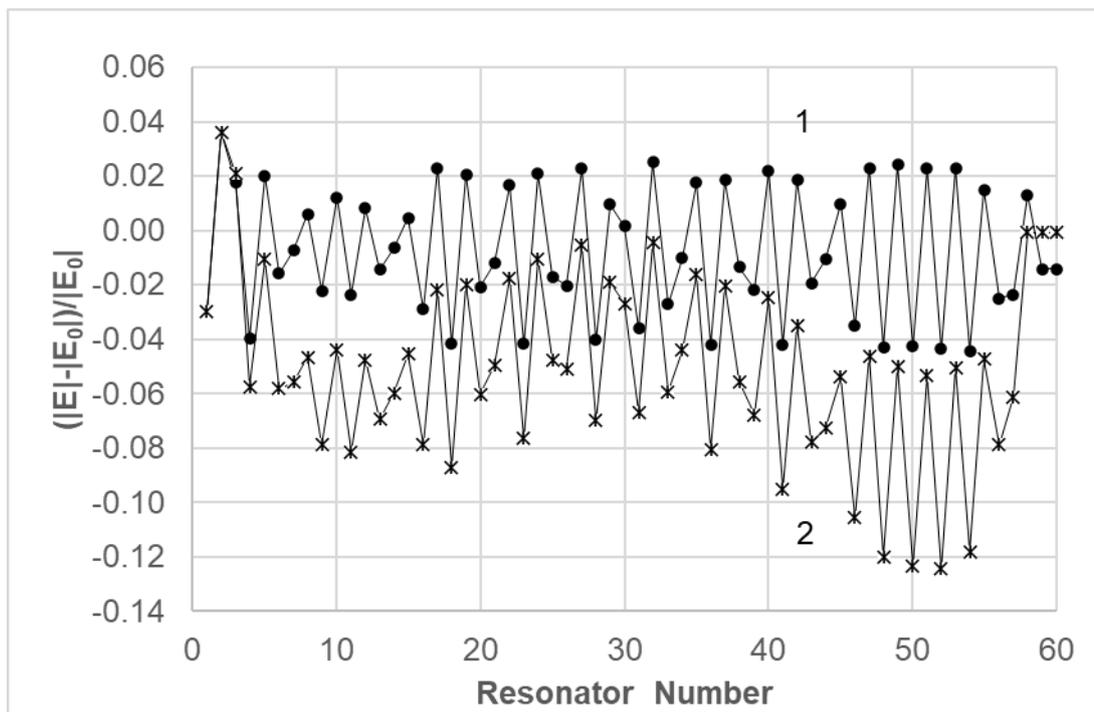

**Figure 4 Comparisons of electric field amplitude distributions calculated on the base of the initial system (39) and WKB approximation (73) - (1), the initial system (39) and Eikonal approximation (72) - (2)**

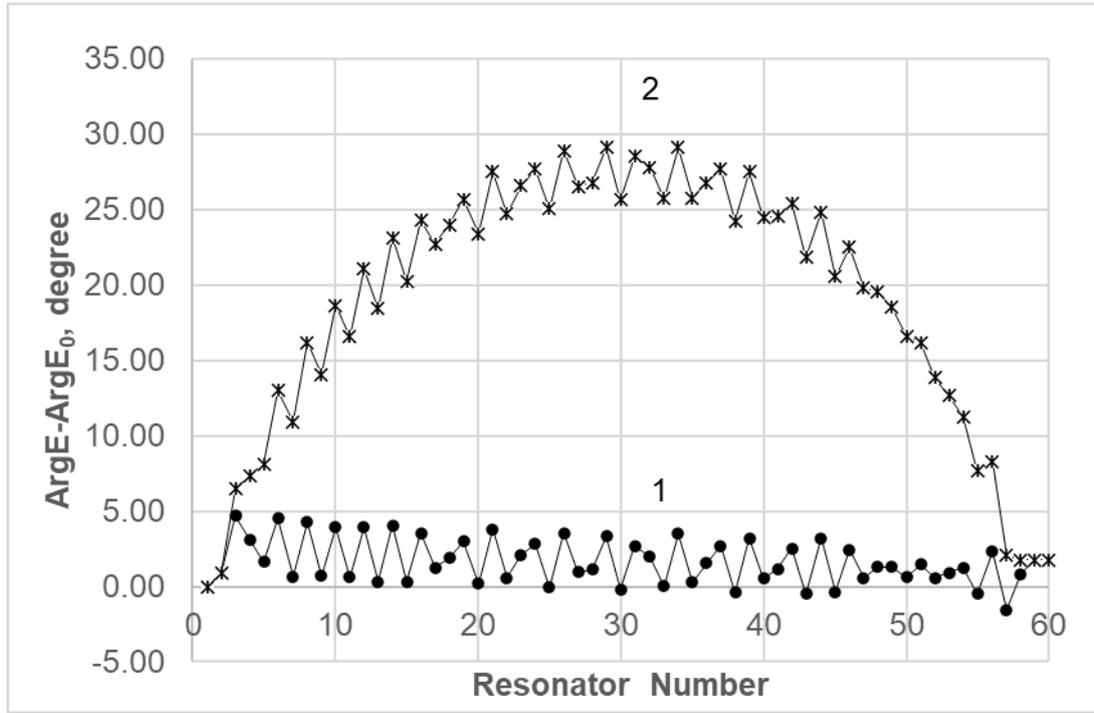

**Figure 5** Comparisons of electric field phase distributions calculated on the base of the initial system (39) and WKB approximation (73) - (1), the initial system (39) and Eikonal approximation (72) - (2)

All above results were calculated on the base of computer codes in which the full systems of linear equations are solved. As in the WKB and Eikonal approximations the matrix equations (72) and (73) have the analytic solutions, we can greatly simplify the system (74)..Under this, it should be borne in mind that one part of the field is described by growing solutions when moving in the positive direction of the waveguide axis; therefore, it is necessary to calculate the field by moving in the negative direction. It is also convenient to use vectors $C^{(k,i)}$ as they directly determine the values of electric field. Such simplified system of equations is

$$C^{(k+1,1)} = \Upsilon^{(k+1,+)} C^{(k,1)}, k = 2,3,...,N_R - 2$$
$$C^{(k,2)} = \Upsilon^{(k,-)} C^{(k+1,2)}, k = N_R - 2,...,3,2, \quad (87)$$

where

$$\Upsilon^{(k+1,+)} = \Xi^{(k+1)} \tilde{\tilde{M}}^{(k+1,1)} \Xi^{(k)-1},$$
$$\Upsilon^{(k,-)} = \Xi^{(k)} \left( \tilde{\tilde{M}}^{(k+1,2)} \right)^{-1} \Xi^{(k+1)-1}. \quad (88)$$

Boundary values of vectors $C^{(k,1)}\left(C^{(2,1)}\right)$ and $C^{(k,2)}\left(C^{(N_R-1,2)}\right)$ are defined by the $4N_m$ equations

$$\left(\varepsilon T^{(1,1)} - W^{(1)}\right) C^{(1)} - \varepsilon T^{(1,3)} \left( C^{(2,1)} + M^{(-)} \Xi^{(N_R-1)-1} C^{(N_R-1,2)} \right) = \frac{b_{w,1}}{a_1 \lambda_1} R_1^{\psi(w,1)},$$

$$-T^{(2,4)} C^{(1)} + \left(T^{(2)} - T^{(2,3)} M^{(3,1)}\right) C^{(2,1)} + \left(T^{(2)} - T^{(2,3)} M^{(3,2)}\right) M^{(-)} \Xi^{(N_R-1)-1} C^{(N_R-1,2)} = 0,$$

$$\left(T^{(N_R)} \Xi^{(N_R)} M^{(N_R-1,1)} - T^{(N_R,4)} \Xi^{(N_R-1)}\right) M^{(+)} C^{(2,1)} + \quad (89)$$

$$+\left(T^{(N_R)} \Xi^{(N_R)} M^{(N_R-1,2)} \Xi^{(N_R-1)-1} - T^{(N_R,4)}\right) C^{(N_R-1,2)} - T^{(N_R,3)} C^{(N_R+1)} = 0,$$

$$\varepsilon T^{(N_R+1,4)} \Xi^{(N_R)} \left( M^{(N_R-1,1)} M^{(+)} C^{(2,1)} + M^{(N_R-1,2)} \Xi^{(N_R-1)-1} C^{(N_R-1,2)} \right) + \left(W^{(2)} - \varepsilon T^{(N_R+1,2)}\right) C^{(N_R+1)} = 0,$$

where

$$M^{(+)} = \prod_{k=3}^{N_R-1} \tilde{\tilde{M}}^{(k,1)},$$
$$M^{(-)} = \prod_{k=N_R-1}^{3} \left(\tilde{\tilde{M}}^{(k,2)}\right)^{-1}. \quad (90)$$

The system of equations (87)-(89) is more suitable for simulation as we have to solve a system of linear equations which dimension is fixed and equals to $4N_m$. This make possible to consider any number of resonators $N_R$. Comparison of results of calculation by using this system and the one based on the solving the full system of linear equations (74) in the WKB approximation shows their good coincidence (error is less than 1.E-6)

As we have already noted, the decomposition (59) is generalization concepts of forward and backward waves, especially in the case of inhomogeneous waveguides. For example, if we consider the case of full reflection from the end of IDLW with zero losses ($a_{N_R+1}=0, \varepsilon=1$) and calculate the electric fields in the centers of resonators that correspond to parts of decompositions $C^{(k,1)}$ and $C^{(k,2)}$ (that are solutions of equations (87))

$$E_z^{(k)}(z_k=d/2, r=0) = E_z^{(k,1)} + E_z^{(k,2)} =$$
$$= \sum_{s=1}^{N_m} T_{s,1}^{E(k)} C_s^{(k,1)} - \sum_{s''=1}^{N_m} T_{s,2}^{E(k)} C_s^{(k+1,1)} + \sum_{s=1}^{N_m} T_{s,1}^{E(k)} C_s^{(k,2)} - \sum_{s''=1}^{N_m} T_{s,2}^{E(k)} C_s^{(k+1,2)}, \quad (91)$$

we find that $E_z^{(k,1)}$ and $E_z^{(k,2)}$ are the complex conjugated values

$$E_z^{(k,1)} = E_z^{(k,2)*}, \quad (92)$$

and their sum is a real value and the distribution $E_z^{(k)}(z_k=d/2, r=0)$ in the case of IDLW is an analog of standing wave. It was shown that in the IDLW with nonzero losses (in our case $\text{Im}\,\varepsilon > 0$) the separation of the electromagnetic field into "forward" and "backward" components is not define uniquely [4]. Indeed, boundary values of vectors $C^{(k,1)}\left(C^{(2,1)}\right)$ and $C^{(k,2)}\left(C^{(N_R-1,2)}\right)$ are defined by parameters of all cells through matrices $M^{(+)}, M^{(-)}$ and reflections from the couplers can affect the boundary values of vectors. It is obvious that for small parameter gradients and losses such influence will be small. To find the value of such influence we made simulation on the base of systems (74) and (87) field distributions for zero and full reflections from the output coupler. Considered section ($N_R=86$, $a_2=1.5$, $a_{85}=1.3$) has such geometrical parameters that the constant gradient distribution is realized under zero reflection from the output coupler. The wall losses were included through the imaginary part of $\varepsilon$: $\text{Im}\,\varepsilon=1.5\text{E-4}$. The working frequency $f=2.856$ GHz, the phase shift per cell $\theta=2\pi/3/$ Amplitudes $\left|E_z^{(k,1)}\right|$ for several cells are presented in Table 11 and Table 12. We can see that for this section the influence of reflections on the distribution of "forward" field is small and we can consider the "forward" $C^{(k,1)}$ and "backward" $C^{(k,2)}$ amplitudes as independent. In addition, the calculation results based on the proposed WKB approach are in good agreement with the exact calculation results. Consequently, the proposed WKB approach can be used to describe sections with small parameter gradients.

**Table 11**

| Cell number | System of equations (74) Amplitudes $\left|E_z^{(k,1)}\right|$ and phases (degree) for two reflection coefficients | | | |
|---|---|---|---|---|
| | $R = 4.11\text{E-}3$ | | $R = 0.59$ | |
| 2 | 1.80858 | 0.0 | 1.81115 | 0.0 |
| 3 | 1.80164 | 120.153 | 1.80401 | 120.153 |
| 4 | 1.80472 | -119.778 | 1.80834 | -119.778 |
| 5 | 1.80467 | 0.489 | 1.80805 | 0.489 |
| | | | | |
| 82 | 1.80586 | -117.205 | 1.80523 | -117.205 |
| 83 | 1.80506 | 2.714 | 1.80443 | 2.714 |
| 84 | 1.80577 | 122.377 | 1.80514 | 122.377 |
| 85 | 1.80585 | -117.857 | 1.80522 | -117.857 |

**Table 12**

| Cell number | System of equations (87) (WKB approach Amplitudes $\left|E_z^{(k,1)}\right|$ and phases (degree) for two reflection coefficients | | | |
|---|---|---|---|---|
| | $R = 4.06\text{E-}3$ | | $R = 0.59$ | |
| 2 | 1.81179 | 0.0 | 1.81115 | 0.0 |
| 3 | 1.80924 | 120.069 | 1.80862 | 120.069 |
| 4 | 1.81114 | -119.648 | 1.81051 | -119.648 |
| 5 | 1.81136 | 0.350 | 1.81073 | 0.350 |
| | | | | |
| 82 | 1.80964 | -117.077 | 1.80901 | -117.077 |
| 83 | 1.80877 | 2.588 | 1.80814 | 2.588 |
| 84 | 1.80953 | 122.504 | 1.80891 | 122.504 |
| 85 | 1.80428 | -117.886 | 1.80365 | -117.886 |

## Conclusions

The presented approach to the description of inhomogeneous disk-loaded waveguides (resonator chains) can be a useful tool in studying the properties of IDLWs. On its basis, various approximate approaches have been developed, including the WKB approximation.